# Periodic Orbit Families in the Gravitational Field of Irregular-shaped Bodies

Yu Jiang[1, 2], Hexi Baoyin[2]

1. State Key Laboratory of Astronautic Dynamics, Xi'an Satellite Control Center, Xi'an 710043, China

2. School of Aerospace Engineering, Tsinghua University, Beijing 100084, China

Y. Jiang (✉) e-mail: jiangyu_xian_china@163.com (corresponding author)

**Abstract**. The discovery of binary and triple asteroids in addition to the execution of space missions to minor celestial bodies in the past several years have focused increasing attention on periodic orbits around irregular-shaped celestial bodies. In the present work, we adopt a polyhedron shape model for providing an accurate representation of irregular-shaped bodies, and employ the model to calculate their corresponding gravitational and effective potentials. We also investigate the characteristics of periodic orbit families and the continuation of periodic orbits. We prove a fact, which provides a conserved quantity that permits restricting the number of periodic orbits in a fixed energy curved surface about an irregular-shaped body. The collisions of Floquet multipliers are maintained during the continuation of periodic orbits around the comet 1P/Halley. Multiple bifurcations in the periodic orbit families about irregular-shaped bodies are also discussed. Three bifurcations in the periodic orbit family have been found around the asteroid 216 Kleopatra, which include two real saddle bifurcations and one period-doubling bifurcation.

## 1. Introduction

The subject of periodic orbits in the gravitational potential of an object is of enduring importance to mathematics, celestial mechanics, and astrodynamics (Galán et al. 2002; Lindner et al. 2010; Chanut et al. 2014). The object itself may be comprised of 1, 2, 3, or more objects (Moore 1993; Najid et al. 2011; Lu et al. 2013, 2014) that include zero-dimensional point masses (Zotos 2015b), one-dimensional massive line segments (Najid et al. 2011), two-dimensional (2-D) massive planar plates (Fukushima 2010), and three-dimensional (3-D) irregular-shaped bodies (Scheeres 2012; Jiang et al. 2015a, b, c). The present study investigates the topological classifications and bifurcations of periodic orbits around irregular-shaped bodies, and is an extension of previous work on the periodic orbits around mass points, line segments, and simple-shaped 3-D bodies.

Several previous studies, such as Moore (1993), Galán et al. (2002), and Zotos (2015b), considered the three body problem, where the three bodies were all considered as point-mass objects. Moore (1993) discussed the braid-type periodic orbits of the three body problem. Galán et al. (2002) presented the stability region for the figure-8 periodic motions of the three-body problem. Zotos (2015b) considered escape orbits in the circular restricted three-body problem, and the orbits were classified into several different cases based on the applied orbital structure.

From the perspective of massive line segments, Riaguas et al. (1999) modeled the motion of bodies in the vicinity of elongated asteroids using a massless point orbiting around a rotating massive line segment. Palacián et al. (2006) investigated the stability of collinear equilibria and the quasi-periodic orbits of a massless particle in the potential of a rotating massive line segment. Lindner et al. (2010) studied the dynamical behaviors of a massive line segment and point mass system, and distinguished several types of orbits, including unstable periodic orbits and stable synchronous orbits. Najid et al. (2011) presented a closed analytical form of the gravitational potential of an inhomogeneous massive line segment, and calculated several orbits around the segment. Blaikie et al. (2014) considered a gravitational system comprised of two massive line segments, and obtained a stable orbit, an unstable orbit, and a chaotic orbit for this dynamic system.

Massive planar plates can include different shapes such as a circular ring, a circular disk, a triangular plate, or a square plate. Broucke & Elipe (2005) derived an expression for the gravitational potential generated by a massive circular ring, and computed several different periodic orbits. Blesa (2006) employed a Poincare section, and respectively calculated the periodic orbits around a massive triangular plate and square plate. Alberti & Vidal (2007) analyzed the gravitational potential of a massive circular disk, and provided the force acting on a massless particle. Fukushima (2010) numerically calculated the gravitational acceleration caused by a massive circular ring or disk, which can be used to calculate the equilibrium points and periodic orbits around them. Zotos (2014, 2015a) investigated the gravitational potential generated by a 2-D perturbed harmonic oscillator, and discussed the escape orbits and chaotic motion of bodies in the potential.

For simple-shaped 3-D bodies, Vasilkova (2005) modeled the gravitational potential of an asteroid using a triaxial ellipsoid. The periodic orbits around equilibrium points of the triaxial ellipsoid were calculated to explain the dynamical behaviors in the vicinity of an asteroid. Chappell et al. (2012) considered the gravitational potential generated by a cube, and modeled its surrounding orbits. Romanov & Doedel (2012) calculated several families of periodic orbits around equilibrium points, and presented a bifurcation diagram of these periodic orbit families. These authors considered their study to be useful for understanding the motion around asteroids 1 Ceres and 433 Eros. Lu et al. (2013, 2014) employed the cellinoid shape model, which is comprised of a collection of eight octants of ellipsoids having different semi-axes, for modeling irregular 3-D shapes and their corresponding gravitational potentials, and calculated the shape parameters and gravitational parameters of cellinoid shapes for 3 Juno and 21 Lutetia.

However, most objects in the solar system deviate considerably from approximations based on point masses, massive line segments, massive planar plates, or simple-shaped 3-D bodies (Ostro et al. 2000; Stooke 2002). Generally, only planets can be modeled as spheroids with small perturbations. Normally, asteroids and comets do not have sufficient mass to generate the required gravitational force for overcoming the solid stress of their constituent materials, so most asteroids and cometary nuclei have irregular shapes (Ostro et al. 2000; Scheeres 2012). These minor celestial bodies not only have non-axisymmetric shapes, but may also have concave and convex shapes. Some asteroids even have highly elongated shapes or contact-binary shapes (Stooke 2002; Neese 2004). A polyhedral shape model comprised of a sufficient number of faces can accurately model irregular and concave shapes, as well as convex surfaces (Werner 1994; Werner and Scheeres 1996; Scheeres 2012; Hirabayashi and Scheeres 2015; Jiang and Baoyin 2014; Jiang 2015; Jiang et al. 2015a, 2015b, 2015c; Chanut et al. 2014, 2015a, 2015b). Therefore, this model can accurately represent any minor celestial body. Using the polyhedral shape model, Scheeres et al. (1996) employed the polyhedral shape model to calculate several periodic orbit families around 4769 Castalia, which has a bifurcated two-lobe irregular shape, and four periodic orbits were obtained near the equilibrium points of the asteroid. Jiang et al. (2014) presented a local dynamical theory regarding the equilibrium points of asteroids, and applied the theory to four asteroids, i.e., 216 Kleopatra, 1620 Geographos, 4769 Castalia, and 6489 Golevka. The researchers distinguished six families of periodic orbits near the equilibrium points of 4769 Castalia. Scheeres (2012) considered the orbital dynamics about the tumbling asteroid 4179 Toutatis, and established the absence of an equilibrium point, and that periodic orbits can only be resonant. Yu & Baoyin (2012) evaluated the orbital dynamics of asteroids using a previously developed theoretical orbit classification in a 4-dimensional symplectic manifold (Marsden & Ratiu 1999), and determined that the gravitational potential of an asteroid presented only seven topological cases of periodic orbits. However, the orbital dynamics in the gravitational potential of an asteroid falls within a 6-dimensional symplectic manifold, the structure of which is different from that of a 4-dimensional symplectic manifold. This structure provides for a total of thirteen different topological cases of periodic orbits in the gravitational potential of an asteroid (Jiang et al. 2015a). Moreover, all four types of bifurcations in the periodic orbit families can be found in the gravitational potential of the primary asteroid of the triple asteroid system 216 Kleopatra (Jiang et al. 2015a). Pseudo period-doubling bifurcation and period-doubling bifurcation of periodic orbit families can coexist about the same asteroids (Jiang et al. 2015b). The topological cases of the periodic orbits vary after period-doubling bifurcation while they remain unchanged after pseudo period-doubling bifurcation. Both bifurcations have Floquet multipliers colliding at −1. Chanut et al. (2014) discussed the minimum radii of equatorial circular orbits, and determined that these radii cannot impact the body of 433 Eros.

For periodic orbits around the equilibrium points of an asteroid, the topological cases of periodic orbits correspond with the topological cases of equilibrium points (Jiang 2015). The calculation of periodic orbits around 101955 Bennu (Jiang et al. 2015b), which has the greatest number of equilibrium points among known asteroids, implies that the resonant periodic orbits around asteroids can also be stable.

Our interest in the dynamical behavior about irregular-shaped bodies mainly concerns the gravitational potential, periodic orbit families, and the multiple bifurcations of periodic orbit families. The goals of the present study are to investigate the topological classifications and bifurcations of periodic orbits about irregular-shaped bodies, and to investigate different types of bifurcations and multiple bifurcations during the continuation of periodic orbits. We computed the gravitational parameters generated by the asteroid 216 Kleopatra and the comet 1P/Halley. The effective potentials and their gradients on different planes are quite different. Furthermore, the shapes of the Hill regions for these two objects are also quite different. We prove a fact, which provides a conserved quantity that permits restricting the number of periodic orbits in a fixed energy curved surface about an irregular-shaped body.

Previous studies have investigated the periodic orbits, periodic orbit families, and bifurcations of periodic orbit families around irregular-shaped bodies, such as 4179 Toutatis (Scheeres 2012), 101955 Bennu (Jiang et al. 2015b), and 216 Kleopatra (Yu and Baoyin 2012; Jiang et al. 2015a, b; Chanut et al. 2015a, 2015b). However, there also exist topological cases that maintain the collisions of Floquet multipliers, which do not lead to the bifurcations of periodic orbit families, and multiple bifurcations of periodic orbit families may also occur. During the continuation of periodic orbits around the comet 1P/Halley, we observe collisions of Floquet multipliers, and topological cases of periodic orbit families maintaining collision during the continuation of periodic orbits. We observe multiple bifurcations, which are comprised of two real saddle bifurcations and a period-doubling bifurcation, during the continuation of periodic orbits about 216 Kleopatra, where the Floquet multipliers can rotate 180 ° in the complex plane during continuation.

**2. Gravitational Field, Floquet multipliers, and Topological Cases of Periodic Orbits**

In this section, we consider the periodical motion of a massless point in the gravitational field of an irregular-shaped body (Baer et al. 2011). The gravitational potential $U$ of the body is calculated using following the polyhedron method (Werner 1994; Werner and Scheeres 1997; Greenberg and Margot 2015):

$$U = \frac{1}{2} G\sigma \sum_{e \in edges} \mathbf{r}_e \cdot \mathbf{E}_e \cdot \mathbf{r}_e \cdot L_e - \frac{1}{2} G\sigma \sum_{f \in faces} \mathbf{r}_f \cdot \mathbf{F}_f \cdot \mathbf{r}_f \cdot \omega_f. \qquad (1)$$

Here, $G = 6.67 \times 10^{-11}$ m$^3$ kg$^{-1}$ s$^{-2}$ is the gravitational constant, $\sigma$ represents the body's bulk density, $\mathbf{r}_e$ and $\mathbf{r}_f$ are body-fixed vectors, where $\mathbf{r}_e$ extends from a field point to any point on an edge $e$, and $\mathbf{r}_f$ extends from a fixed point to any point on a face $f$, $\mathbf{E}_e$ and $\mathbf{F}_f$ are body-fixed tensors, where $\mathbf{E}_e$ represents the geometric parameters of $e$, and $\mathbf{F}_f$ represents the geometric parameters of $f$, $L_e$ represents the integration factor that operates between a fixed point and $e$, and $\omega_f$ represents the solid angle relative to the fixed point.

We now consider the equation of motion of a massless particle in the gravitational field of an irregular-shaped body in Cartesian coordinates x, y, and z. We denote the body's rotational angular velocity as $\boldsymbol{\omega}$, the body-fixed vector from the asteroid's center of mass to the particle as $\mathbf{r}$, and $\omega$ as the norm of $\boldsymbol{\omega}$. We define the unit vector $\mathbf{e_z}$ of the z-axis for the asteroid's body-fixed frame of reference as $\boldsymbol{\omega} = \omega \mathbf{e_z}$. Based on these definitions, the dynamical equation of the particle can be written as follows (Jiang and Baoyin 2014).

$$\begin{cases} \ddot{x} - \dot{\omega}y - 2\omega\dot{y} - \omega^2 x + \frac{\partial U}{\partial x} = 0 \\ \ddot{y} + \dot{\omega}x + 2\omega\dot{x} - \omega^2 y + \frac{\partial U}{\partial y} = 0 \\ \ddot{z} + \frac{\partial U}{\partial z} = 0 \end{cases} \qquad (2)$$

The asteroid rotates uniformly, so the Jacobi integral $H$ is given as

$$H = U + \frac{1}{2}\left(\dot{x}^2 + \dot{y}^2 + \dot{z}^2\right) - \frac{\omega^2}{2}\left(x^2 + y^2\right). \qquad (3)$$

The effective potential $V$ is defined as

$$V = U - \frac{\omega^2}{2}\left(x^2 + y^2\right). \qquad (4)$$

We now calculate $V$ for 216 Kleopatra. The lengths of the three axes of 216 Kleopatra are 217, 94, and 81 km, and the density and rotational periods are 3.6 g·cm$^{-3}$ and 5.385 h, respectively (Ostro et al. 2000; Descamps et al. 2011). The shape and gravitational field of 216 Kleopatra are generated by the polyhedral model (Werner and Scheeres 1997). The modeling of 216 Kleopatra employed 2,048 vertices and 4,096 faces (Neese 2004). Figure 1 shows a 3-D contour plot of $V$ for 216 Kleopatra, where the values for $V$ are respectively plotted in the xy, yz, and zx planes. The figure indicates that the gradient of $V$ in the yz plane is quite different from those in the xy and zx planes. Asteroid 216 Kleopatra has seven equilibrium points (Chanut et al. 2015a; Jiang et al. 2015b; Jiang et al. 2016), all of which are out-of-plane equilibrium points; however, the positions of these equilibrium points are near the xy plane because the z axis serves as the rotation axis of the asteroid (Jiang et al. 2015b). Figure 1 also shows that the Hill region of 216 Kleopatra has a bifurcated two-lobe structure, which is caused by the contact-binary shape of the asteroid.

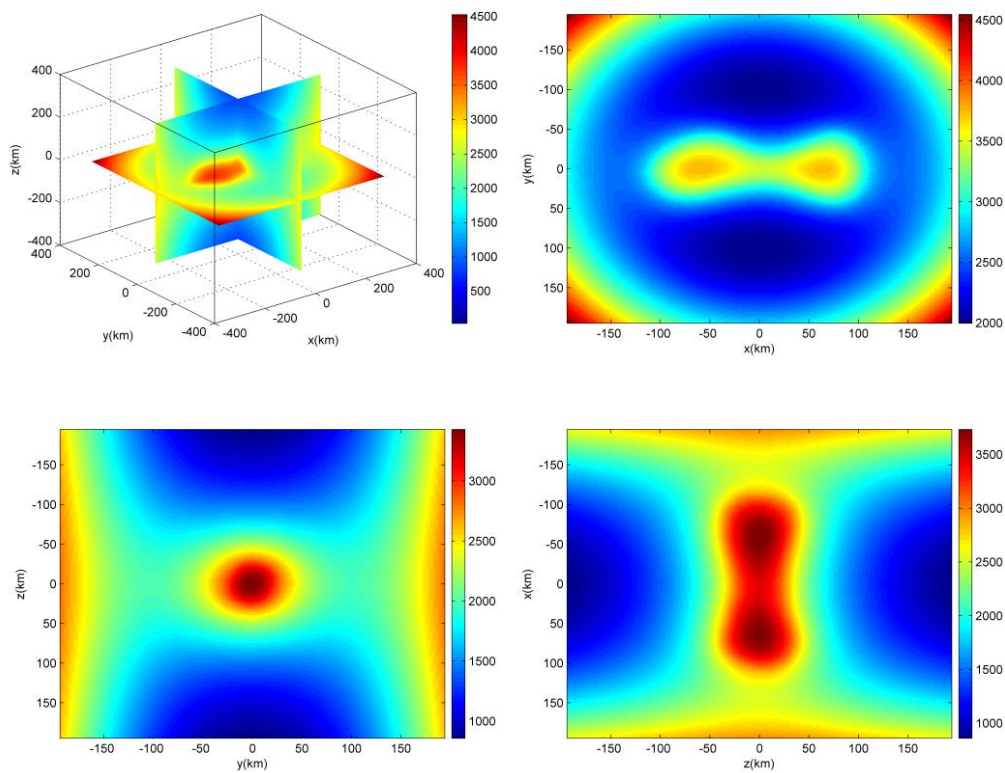

Figure 1. A 3-D contour plot of the effective potential for 216 Kleopatra (unit: $m^2 \cdot s^{-2}$).

We now calculate $V$ for 1P/Halley. The lengths of the three axes of 1P/Halley are 16.831, 8.7674, and 7.7692 km, and the density and rotational periods are 0.6 $g \cdot cm^{-3}$ and 52.8 h, respectively (Sagdeev et al. 1986, 1988; Peale & Lissauer 1989). The shape and gravitational potential of 1P/Halley were generated by the polyhedral model (Werner & Scheeres 1997), and the modeling employed 2,522 vertices and 5,040 faces (Stooke 2002). Figure 2 shows a 3-D contour plot of $V$ for 1P/Halley, where the values of *V* are respectively plotted in the xy, yz, and zx planes. The figure indicates that the values of *V* at different positions located at an equivalent distance from the center of mass of the comet may be quite different. Additionally, the gradients of *V* in the xy and zx planes are more irregular than that in the yz plane. Figure 2 indicates that the shape of the Hill region for 1P/Halley is a non-axisymmetric structure, which is different from that for 216 Kleopatra.

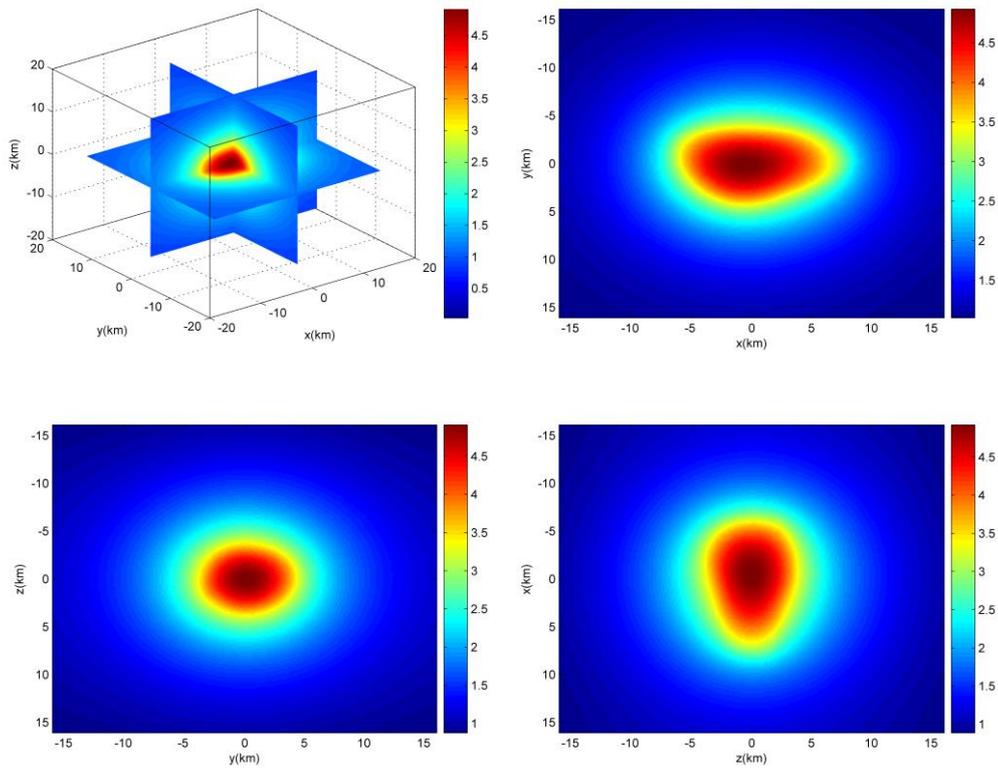

Figure 2. A 3-D contour plot of the effective potential for 1P/Halley (unit: $m^2 \cdot s^{-2}$).

We now consider periodic orbits about irregular-shaped bodies. A periodic orbit in the gravitational field of an irregular-shaped body has a monodromy matrix $M = \Phi(T)$, where $\Phi(\cdot)$ is the state transition matrix. If we denote $p$ as a periodic orbit, then $\Phi(\cdot)$ can be calculated as

$$\Phi(t) = \int_0^t \frac{\partial \mathbf{f}}{\partial \mathbf{z}}(p(\tau)) d\tau. \tag{5}$$

Here, $\mathbf{z}(t) = \mathbf{f}(t, \mathbf{z}_0)$ is the solution of Eq. (2) that satisfies $\mathbf{f}(0, \mathbf{z}_0) = \mathbf{z}_0$, where $t$ represents time. If we let $\mathbf{f}^t : \mathbf{z} \to \mathbf{f}(t, \mathbf{z})$, then $\mathbf{f}^t$ represents the orbit of the particle relative to the asteroid. Floquet multipliers (Jiang et al. 2015a) of a periodic orbit have the form 1, $-1$, $\text{sgn}(\alpha) e^{\pm \alpha}$ $(\alpha \in \mathrm{R}, |\alpha| \in (0,1))$, $e^{\pm i\beta}$ $(\beta \in (0, \pi))$, and $e^{\pm \sigma \pm i\tau}$ $(\sigma, \tau \in \mathrm{R};\ \sigma > 0, \tau \in (0, \pi))$, where $\text{sgn}(\alpha) = \begin{cases} 1 & (\text{if } \alpha > 0) \\ -1 & (\text{if } \alpha < 0) \end{cases}$. Here, periodic orbits have at least two Floquet multipliers equal to +1.

The periodic orbits around irregular-shaped bodies have seven normal topological cases, where each case has exactly two Floquet multipliers equal to 1. Table A1 in the Appendix presents the topological classifications of periodic orbits around irregular-shaped bodies. If the algebraic multiplicity of Floquet multipliers equal to +1 is 2, then the periodic orbit is denoted as non-degenerate (Meyer 1998); otherwise, a periodic orbit is denoted as a degenerate periodic orbit if and only if it has at least four Floquet multipliers equal 1. The Krein collision cases of periodic orbits include a single case only, Case K1: $\gamma_j (\gamma_j = 1; j = 1, 2)$ and $e^{\pm i\beta_j} (\beta_j \in (0, \pi), \beta_1 = \beta_2; j = 1, 2)$. The degenerate periodic cases of periodic orbits have four cases, denoted as Case DP1: $\gamma_j (\gamma_j = 1; j = 1, 2, 3, 4)$ and $e^{\pm i\beta} (\beta \in (0, \pi))$; Case DP2: $\gamma_j (\gamma_j = 1; j = 1, 2, 3, 4)$ and $e^{\pm \alpha} (\alpha \in \mathrm{R}, 0 < \alpha < 1)$; Case DP3: $\gamma_j (\gamma_j = 1; j = 1, 2, 3, 4)$ and $e^{\pm \alpha} (\alpha \in \mathrm{R}, -1 < \alpha < 0)$; and Case DP4: $\gamma_j (\gamma_j = 1; j = 1, 2, \cdots, 6)$. The degenerate real saddle cases of periodic orbits have two cases, denoted as Case DR1: $\gamma_j \left( \gamma_j = 1; j = 1,2 \right)$ and $e^{\pm \alpha_j} \left( \alpha_1, \alpha_1 \in \mathrm{R}; 0 < \alpha_1 = \alpha_2 < 1 \right)$; and Case DR2: $\gamma_j \left( \gamma_j = 1; j = 1,2 \right)$ and $e^{\pm \alpha_j} \left( \alpha_1, \alpha_1 \in \mathrm{R}; -1 < \alpha_1 = \alpha_2 < 0 \right)$. The period-doubling cases of periodic orbits have five cases, denoted as Case PD1: $\gamma_j \left( \gamma_j = 1; j = 1,2,3,4 \right)$ and $\gamma_j \left( \gamma_j = -1; j = 5,6 \right)$; Case PD2: $\gamma_j \left( \gamma_j = 1; j = 1,2 \right)$ and $\gamma_j \left( \gamma_j = -1; j = 3,4,5,6 \right)$; Case PD3: $\gamma_j \left( \gamma_j = 1; j = 1,2 \right)$, $\gamma_j \left( \gamma_j = -1; j = 3,4 \right)$ and $e^{\pm i\beta} \left( \beta \in (0, \pi) \right)$; Case PD4: $\gamma_j \left( \gamma_j = 1; j = 1,2 \right)$, $\gamma_j \left( \gamma_j = -1; j = 3,4 \right)$ and $e^{\pm \alpha} \left( \alpha \in \mathrm{R}, 0 < \alpha < 1 \right)$; and Case PD5: $\gamma_j \left( \gamma_j = 1; j = 1,2 \right)$, $\gamma_j \left( \gamma_j = -1; j = 3,4 \right)$ and $e^{\pm \alpha} \left( \alpha \in \mathrm{R}, -1 < \alpha < 0 \right)$.

The topological classification of periodic orbits is based on the distribution of Floquet multipliers of periodic orbits in the complex plane. Bifurcations may occur when the periodic orbits are continued. A period-doubling bifurcation occurs when two Floquet multipliers equal $-1$. A Neimark-Sacker bifurcation occurs when two Floquet multipliers collide on the unit circle, where the collision in this case is denoted as a Krein collision. A real saddle bifurcation occurs when two Floquet multipliers collide on the real axis. Considering the conjugation of the Floquet multipliers, the number of collisions always equals 2. The seven normal topological cases of periodic orbits have no degenerate period, period-doubling, or collision. In the previously discussed work of Scheeres (2012), several periodic orbits were calculated around the tumbling asteroid 4179 Toutatis with periods equal to one or two times the rotational period in the body-fixed frame. Jiang (2015) presented several periodic orbit families around the relative equilibria of 216 Kleopatra, and found that the topological classification of these periodic orbit families belongs only to Case N1 or Case N6. Here, the topological classification of the periodic orbit families around the relative equilibria near the long axis of the asteroid conforms to Case N6, while the topological classification of the periodic orbit families around the relative equilibria near the short axis of the asteroid conforms to Case N1.

We now consider periodic orbit families about irregular-shaped bodies. The following fact describes the Floquet multipliers and the number of periodic orbits and periodic orbit families about irregular-shaped bodies.

**Fact 1.** Suppose the existence of $N$ periodic orbits in the energy curved surface $H = C$ in the gravitational potential of an irregular-shaped body, and denote $p_i$ as the $i$th periodic orbit and $\varsigma_j(p_i)$ as the $j$th Floquet multiplier of $p_i$, $\varsigma_5(p_i) = \varsigma_6(p_i) = 1$. We then have $\sum_{i=1}^{N}\left[\operatorname{sgn}\prod_{j=1}^{4}\varsigma_j(E_i)\right] = const$.

**Proof:**

Consider a Poincaré section of the periodic orbits on $H = C$. The Poincaré section is a 4-dimensional subspace. Let $\varphi(\mathbf{z}) = \int \Phi(t) d\mathbf{z}$, where $\varphi(\mathbf{z})$ is the primitive function of $\Phi(t)$, which yields $\nabla\varphi(\mathbf{z}) = \Phi(t)$. Then, the periodic orbit is the fixed point of $\varphi(\mathbf{z})$.

Let $f(\mathbf{z})$ be given as

$$f(\mathbf{z}) = \ln\left(\frac{\varphi(\mathbf{z},t)}{\varphi(\mathbf{z},0)}\right). \tag{6}$$

Then, the periodic orbit is the zero point of $f(\mathbf{z})$. Let A be the open set which includes all periodic orbits on $H = C$.

Applying the topological degree theory to the function $f(\mathbf{z})$ yields

$$\deg(f,\ A,\ 0) = \sum_{i=1}^{N}\left[\operatorname{sgn}(\det(\ln M_i))\right] = \sum_{i=1}^{N}\left[\operatorname{sgn}\prod_{j=1}^{4}(\ln\varsigma_i(p_k))\right] = const. \tag{7}$$

□

This fact provides a conserved quantity that can restrict the number of periodic orbits in $H = C$ for the gravitational potential of an irregular-shaped body. If the parameter $C$ varies, then the conserved quantity can restrict the number of periodic orbit families in the gravitational potential of an irregular-shaped body. For a periodic orbit $p_k$, we denote the index by $ind(p_k) = \prod_{j=1}^{4}(\ln \varsigma_i(p_k))$. If $ind(p_k) = 0$, $p_k$ is denoted as degenerate, otherwise $p_k$ is denoted as non-degenerate. For the period-doubling case, Case PD1, we also have $ind(p_k) = 0$. For the two normal topological cases, Case N6 and Case N7, as well as the period-doubling cases, Case PD4 and Case PD5, $ind(p_k) = -1$. For other topological cases of periodic orbits, $ind(p_k) = 1$.

We obtain the following facts from Fact 1.

**Fact 2.** The number of non-degenerate periodic orbits in the gravitational field of an irregular-shaped body varies in pairs during the continuation of the orbits.

**Fact 3.** During continuation, the degenerate periodic orbit in the gravitational field of an irregular-shaped body will change to $k(k \in Z, k \geq 0)$ degenerate periodic orbits and $2l(l \in Z, l \geq 0)$ non-degenerate periodic orbits, where $Z$ is the set of all integers. Here, if $k = l = 0$, the degenerate periodic orbit will disappear.

**Fact 4.** For $p_k$, suppose that the number of Floquet multipliers that lie on the real axis and outside the unit circle is $m(p_k)$. Then, $\sum_{k=1}^{N}(-1)^{m(p_k)} = const$. Assume that the number of Floquet multipliers that lie on the real axis and inside the unit circle is $n(p_k)$; then, $m(p_k) = n(p_k)$. Thus, $\sum_{k=1}^{N}(-1)^{n(p_k)} = const$.

**Fact 5.** The conserved quantity of the periodic orbits is 1, which yields $\sum_{i=1}^{N}\left[\operatorname{sgn}\prod_{j=1}^{4}\left(\ln\varsigma_i(p_k)\right)\right] = 1$. Thus, the number of non-degenerate periodic orbits in the gravitational field of an irregular-shaped body is an odd number. The number of non-degenerate periodic orbits can only vary between odd numbers during the continuation of a periodic orbit.

## 3. Bifurcations of Periodic Orbit Families

The periodic orbit families about irregular-shaped bodies include two types of bifurcations (Broucke and Elipe 2005; Romanov and Doedel 2014; Jiang et al. 2015a, 2015b). The first type of bifurcation of periodic orbit families is represented by the variation in the topological classification of the periodic orbits during their continuation, which includes 4 types, i.e., tangent bifurcations, period-doubling bifurcations, Neimark-Sacker bifurcations, and real saddle bifurcations (Jiang et al. 2015a). As for the variety of topological classifications of the periodic orbits, if two Floquet multipliers cross -1, a period-doubling bifurcation occurs; if two Floquet multipliers cross 1, a tangent bifurcation occurs; Neimark-Sacker bifurcations appear when the periodic orbit has two equal Floquet multipliers on the unit circle away from {-1,1}, real saddle bifurcations appear when the periodic orbit has two equal Floquet multipliers on the x-axis away from {-1,1}.

The second type of bifurcation of periodic orbit families is represented by the generation or annihilation of periodic orbit families. When the parameters (for instance, the rotation speed, the shape of the irregular-shaped bodies, the Jacobian integral, the period, or the natural parameter) are varied, one periodic orbit family may change to two or more periodic orbit families, or more than one periodic orbit family may become one periodic orbit family. In Jiang et al. (2015c), we find collision and annihilation of equilibrium points in the potential of asteroids after a variation of parameters. Equilibrium points with eigenvalues on the imaginary axis have emanating periodic orbit families (Scheeres et al. 1996; Scheeres 2012; Jiang et al. 2014; Jiang 2015). If the equilibrium point is linear stable, then three families of periodic orbits have been shown to exist around the equilibrium point. If the equilibrium point has exactly two pairs of eigenvalues on the imaginary axis, two families of periodic orbits have been shown to exist around the equilibrium point. If the equilibrium point has only one pair of eigenvalues on the imaginary axis, only one family of periodic orbits has been shown to exist around the equilibrium point (Jiang 2015). When varying parameters (for instance, the rotation speed, the shape of the irregular bodies), if two equilibrium points collide and annihilate each other, then the periodic orbit families that correspond to the equilibrium point also collide and disappear (For a rotating homogeneous cube, there is no collision and annihilation of equilibrium points during the variety of rotating speed.). This is the second type of bifurcation of periodic orbit families around irregular bodies. The same applies to the generation and separation of equilibrium points and periodic orbit families around equilibrium points.

Using the principle of least action, we can conclude that the periodic orbit in Eq. (2) which has a fixed Hamiltonian function can be translated into an equilibrium point of another system.

The Lagrange function (Jiang and Baoyin 2014) can be written as

$$L = \frac{1}{2}(\dot{x}^2 + \dot{y}^2 + \dot{z}^2) + \frac{1}{2}\omega^2(x^2 + y^2) + \omega(x\dot{y} - \dot{x}y) - U. \tag{8}$$

The Jacobi integral $H$ is the Hamilton function (see Eq.(3)).

Then Eq. (2) can be written as

$$\frac{d}{dt}\left(\frac{\partial L}{\partial \dot{\mathbf{q}}}\right) = \frac{\partial L}{\partial \mathbf{q}} \tag{9}$$

or

$$\begin{cases} \dot{\mathbf{p}} = -\dfrac{\partial H}{\partial \mathbf{q}} \\ \dot{\mathbf{q}} = \dfrac{\partial H}{\partial \mathbf{p}} \end{cases}. \tag{10}$$

where $\mathbf{p} = (\dot{\mathbf{r}} + \boldsymbol{\omega} \times \mathbf{r})$ and $\mathbf{q} = \mathbf{r}$.

Let the period of the periodic orbit be $T$. Define the functional operation as

$$\varphi(\mathbf{q}) = \int_0^T L(\mathbf{q}(t), \dot{\mathbf{q}}(t), t) dt. \tag{11}$$

Then the Euler equation of this functional operation is Eq. (9).

Define

$$\psi(\mathbf{p}, \mathbf{q}) = \int_0^T \left[\mathbf{p}(t) \cdot \dot{\mathbf{q}}(t) - H(\mathbf{p}(t), \mathbf{q}(t), t)\right] dt. \tag{12}$$

The above functional operation is obtained by using $\mathbf{p} \cdot \dot{\mathbf{q}} - H(\mathbf{p}, \mathbf{q}, t)$ to replace $L(\mathbf{q}, \dot{\mathbf{q}}, t)$. The Euler equation of the above functional operation is Eq. (10).

Let $\mathbf{J}$ be a standard symplectic matrix

$$\mathbf{J} = \begin{bmatrix} \mathbf{0}_n & -\mathbf{I}_n \\ \mathbf{I}_n & \mathbf{0}_n \end{bmatrix}, \quad \mathbf{J}^2 = -\mathbf{I}_{2n}. \tag{13}$$

$(\mathbf{J}\mathbf{u}, \mathbf{v}) = -(\mathbf{u}, \mathbf{J}\mathbf{v})$ is satisfied for all $\mathbf{u}, \mathbf{v} \in R^{2n}$.

Let $\mathbf{u} = (\mathbf{p}, \mathbf{q})^T$, then Eq. (10) can be written as

$$\dot{u}(t) = J\nabla H(t, u(t)) \tag{14}$$

or

$$J\dot{u}(t) + \nabla H(t, u(t)) = 0. \tag{15}$$

So

$$\begin{cases} J\dot{u}(t) = (-\dot{q}, -\dot{p})^T \\ (J\dot{u}(t)) \cdot u(t) = p \cdot (-\dot{q}) + \dot{p} \cdot q \end{cases}. \tag{16}$$

Because the period is $T$, then

$$\int_0^T (p(t) \cdot \dot{q}(t)) dt = \frac{1}{2} \int_0^T \left\{ p(t) \cdot \dot{q}(t) + \frac{d}{dt}[p(t) \cdot q(t)] - [\dot{p}(t) \cdot q(t)] \right\} dt$$
$$= -\frac{1}{2} \int_0^T \{\dot{p}(t) \cdot q(t) + p(t) \cdot [-\dot{q}(t)]\} dt = -\frac{1}{2} \int_0^T [J\dot{u}(t) \cdot u(t)] dt \tag{17}$$

If we denote $a \cdot b = (a, b)$, then

$$\int_0^T (p(t), \dot{q}(t)) dt = -\frac{1}{2} \int_0^T [J\dot{u}(t), u(t)] dt. \tag{18}$$

So the functional operation $\psi$ can be given by

$$\psi(u) = -\int_0^T \left[ \frac{1}{2}(J\dot{u}(t), u(t)) + H(t, u(t)) \right] dt. \tag{19}$$

Thus the critical point of $\psi$ is the solution of the Hamiltonian system with the two boundary condition $u(0) = u(T)$, in other words, the equilibrium point of Eq. (19) and the periodic orbit of Eq. (10) with period $T$ are equivalent. That is, the periodic orbits with period T of Eq. (2) and the equilibrium points of functional operation $\psi(u)$ are equivalent. Because the equilibrium points may collide and annihilate each other during the variation of parameters, the periodic orbits may also collide and annihilate each other (Romanov & Doedel 2012, 2014) during the variation of parameters (for instance, the Jacobian integral, the period, or the natural parameter).

The difference between these two types of periodic orbit family bifurcations is that the first type falls within the periodic orbit families while the second type falls between the periodic orbit families. There is no generation or annihilation of periodic orbit families in the first type of bifurcation, and the topological classification of the periodic orbits varies when the periodic orbits are continued and conform to a periodic orbit family. Here, we analyze only the bifurcations and pseudo bifurcations within periodic orbit families.

Bifurcations may occur during the continuation of periodic orbits. Previous studies have considered single bifurcations within periodic orbit families about irregular-shaped bodies (Broucke & Elipe 2005; Jiang et al. 2015a, b). However, as discussed above, multiple bifurcations are possible during the continuation of periodic orbits, and, even if these bifurcations involve the same type of bifurcation, the transition pathways may be different. In addition, different types of bifurcations and pseudo bifurcations can occur in the same periodic orbit family.

Jiang et al. (2015b) presented different scenarios for sequences of period-doubling bifurcations in a periodic orbit family. However, real saddle bifurcations and tangent bifurcations also include several different transition pathways. Table A2 in the Appendix lists the transition pathways for these four bifurcations. The opposite pathway corresponds to an equivalent bifurcation sequence, and is therefore omitted in the Table. For example, period-doubling bifurcation I is represented by the transitions Case DP1→Case PD1→Case DP3 or Case DP3→Case PD1→Case DP1, which, because they are simply opposite pathways, only the first of the two possible sequences is listed. Neimark-Sacker bifurcations have only a single pathway, i.e., Case N1→Case PK1→Case N2 or its opposite. During the continuation of periodic orbits, the sojourn time of periodic orbits at the bifurcations is not necessarily zero. If the sojourn time is zero, there is only one periodic orbit at the bifurcations; otherwise, there are infinitely many periodic orbits at the bifurcations. Taking the Neimark-Sacker bifurcation as an example, if the sojourn time is not zero, then several periodic orbits can conform to topological Case PK1 during the periodic orbit continuation. Furthermore, the Neimark-Sacker bifurcation may not occur during the continuation, but, rather, the Floquet multipliers may collide on the unit circle, and the collision can be sustained along part of a periodic solution family. This is the collision maintaining attribute of the topological cases that leads to pseudo Neimark-Sacker bifurcation.

With regard to the occurrence of multiple bifurcations during the continuation of periodic orbits, the rules of multiple bifurcations follow the combinations of single bifurcations presented in Table A2. For example, the pathway Case N1 $\to$ Case DR2 $\to$ Case N5 $\to$ Case PD5 $\to$ Case N7 leads to a real saddle bifurcation and a period-doubling bifurcation, which are binary bifurcations. Triple bifurcations and quadruple bifurcations also have several pathways such as Case N5 $\to$Case DR2$\to$Case N1$\to$Case PK1$\to$Case N2$\to$Case PD3$\to$Case N7$\to$Case DP3$\to$Case N4. These pathways include all four bifurcations, and the sequence is real saddle, Neimark-Sacker, period-doubling, and tangent bifurcations. The index of periodic orbits varies from +1 to −1 to 0 to +1, where the index equals −1 when the periodic orbit conforms to Case N7, and it equals 0 when the periodic orbit conforms to Case DP3.

We now continue the periodic orbit families around 216 Kleopatra and 1P/Halley. The dynamical mechanism around irregular-shaped celestial bodies confirms the possible continuation phenomena. The continuation procedure is based on the optimal method to calculate the step size in the gradient direction (Muñoz-Almaraz et al. 2003). The dynamical mechanism around irregular celestial bodies confirms the possible continuation phenomena. The continuation method (Muñoz-Almaraz et al. 2003) is

$$\mathbf{X}_{i+1} = \mathbf{X}_i + \varepsilon \cdot \delta \mathbf{X}_i, \tag{8}$$

where the vector $\mathbf{X}_i$ represents the position and velocity of the particle and $\delta \mathbf{X}_i$ represents the gradient direction of the $i$-th iteration, and $\varepsilon$ is the step size. In addition, the following definitions (Werner & Scheeres 1997; Jiang et al. 2015b) are applied.

$$\begin{cases} \delta \mathbf{X}_i = \begin{pmatrix} \boldsymbol{\omega} \times (\boldsymbol{\omega} \times \mathbf{r}_i) + \nabla U \\ \mathbf{v}_i \end{pmatrix} \\ \nabla U = -G\sigma \sum_{e \in edges} \mathbf{E}_e \cdot \mathbf{r}_e \cdot L_e + G\sigma \sum_{f \in faces} \mathbf{F}_f \cdot \mathbf{r}_f \cdot \omega_f \end{cases} \tag{9}$$

Here, $\nabla U$ represents the gradient of the potential. Collision with the irregular-shaped body ends the orbit continuation procedure.

Prior to continuing the periodic orbit families, a great many single periodic orbits are searched randomly based upon the fixing of a Poincaré section. The error limit is set to be $1.0 \times 10^{-6}$. Among the 10,000 orbits calculated, no more than approximately 60 are selected. Equation (5) is employed to calculate $M$ for each orbit, and the eigenvalues of $M$ determine the Floquet multipliers, which are then employed to select those orbits to be continued for obtaining the periodic orbit families. Some of the periodic orbits conform to similar geometrical shapes and equivalent topological cases, and then the continuation of these periodic orbits leads to a single periodic orbit family for orbit continuation.

If Floquet multipliers exist outside the unit circle, and, in particular, if at least one Floquet multiplier has a very large norm, for instance, 100, the periodic orbit will be unstable, and converge with difficulty. Under this condition, the value of $\varepsilon$ is reduced. The value of $\varepsilon$ is also reduced if bifurcation occurs to determine the bifurcation position.

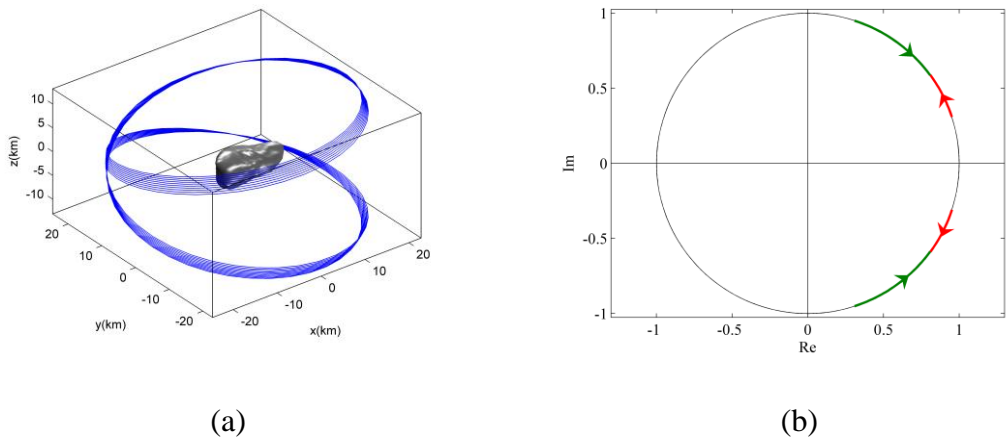

(a)                                      (b)

Figure 3. The continuation of periodic orbits about 1P/Halley (a), and the corresponding collision of Floquet multipliers (b).

Figures 3(a) and (b) respectively illustrate the continuation of periodic orbits about 1P/Halley, and the corresponding collision of Floquet multipliers. In Figure 3(a), we plotted 19 different periodic orbits during the continuation, and Figure 3(b) shows the changes in the Floquet multipliers of these periodic orbits during the continuation. This family of periodic orbits is stable, and the collision of Floquet multipliers is maintained, leaving the topological case of the periodic orbits unchanged. The topological cases for these periodic orbits follow the transition Case 2 → Case K1 → Case 2, which belong to the pseudo Neimark-Sacker bifurcation of periodic orbits. Because of the maintenance of collision, the sojourn time of Case K1 for the periodic orbits is not zero. Theoretically, there are infinite periodic orbits belonging to Case K1 in the continuation of these periodic orbits. Using numerical methods to continue the periodic orbits, we find that there are several periodic orbits that correspond to Case K1. The numerical method we employed does not allow for the determination of orbits of infinite period. The ratios of these continued orbit periods to the period of the comet fall within the interval [1.01193, 1.01243], which indicates that the periodic orbit family is approximately 1:1 resonant. The result implies that the Krein collision of Floquet multipliers may not lead to Neimark-Sacker bifurcation of the periodic orbit family about irregular-shaped bodies, and, after the Krein collision, the Floquet multipliers may not separate immediately. As such, the Krein collision may persist during the continuation.

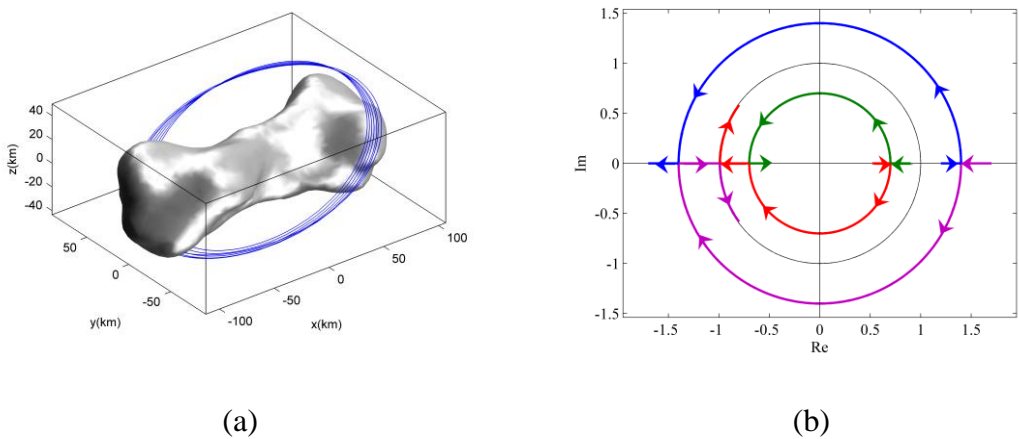

(a)                                   (b)

Figure 4. The continuation of periodic orbits about 216 Kleopatra (a), and the binary real saddle bifurcation as well as the period-doubling bifurcation (b).

Figures 4(a) and (b) respectively present the continuation of periodic orbits about 216 Kleopatra, and the binary real saddle bifurcation as well as the period-doubling bifurcation. Using the continuation procedure, thousands of periodic orbits were computed. Six different topological cases were encountered during the continuation, and Figure 4(a) illustrates the 6 different periodic orbits with different topological cases encountered. Figure 4(b) shows the corresponding change in the Floquet multipliers of these periodic orbits during the continuation. Figure 4(b) indicates that two Floquet multipliers rotate outside the unit circle and two rotate inside the unit circle during the multiple bifurcations. The rotation trajectories of the four Floquet multipliers are not strictly circular, which indicates that the norms of the Floquet multipliers vary during the continuation. The ratio of the period of this orbit family to the period of the asteroid is not a simple integer ratio, which implies that the periodic orbit family is not resonant. During the continuation, the periods of these orbits vary from 2.4257 h to 2.3855 h. The topological cases for these periodic orbits follow the transition Case N3→Case DR1→Case N1→Case DR2→Case N5→Case PD5→Case N7, which includes three bifurcations during the continuation. The index of the periodic orbits varies from +1 to −1 to +1, where the index equals −1 when the periodic orbits conform to Case PD5.

The first bifurcation is the real saddle bifurcation with the pathway Case N3 → Case DR1 → Case N1. Before the first bifurcation, the topological classification of the periodic orbit conforms to Case N3 with four Floquet multipliers on the real axis. During the continuation, two pairs of Floquet multipliers on the real axis approach each other along the real axis and collide, transitioning the topological classification of the periodic orbit to Case DR1. After the collision, four Floquet multipliers leave the real axis and enter the complex plane, transitioning the topological classification of the periodic orbit to Case N1. Further continuation of the periodic orbits shows that these four complex Floquet multipliers rotate 180°, and reach the second bifurcation. To visualize the dynamical behaviors on the Poincaré section of the periodic orbit, note that the rotation of Floquet multipliers not only means that the intersection point of the periodic orbit on the Poincaré section rotates 180°, but also that the orbits around the periodic orbit rotate 180° such that the intersection point of the orbit on the Poincaré section rotates 180°. In addition, the submanifold of the periodic orbits also rotates 180°.

The second bifurcation is also a real saddle bifurcation given by the pathway Case N1 $\rightarrow$ Case DR2 $\rightarrow$ Case N5. After the rotation of the Floquet multipliers in the complex plane, the Floquet multipliers are in the negative half-plane, i.e., the real parts of these four Floquet multipliers are less than zero. Then, two pairs of Floquet multipliers in the complex plane approach each other and collide on the real axis, transitioning the topological classification of the periodic orbit to Case DR2. After collision, four Floquet multipliers remain on the real axis, and two of them approach −1 and collide, transitioning the topological classification of the periodic orbit to Case N5.

The third bifurcation is the period-doubling bifurcation given by the pathway Case N5 $\rightarrow$ Case PD5 $\rightarrow$ Case N6. At the beginning in Case N5, four Floquet multipliers are on the real axis, i.e., the −x axis. Two of the Floquet multipliers, one greater than −1 and the other less than −1, approach −1 and collide, transitioning the topological classification of the periodic orbit to Case PD5 with the inception of period-doubling bifurcation. After collision, these two Floquet multipliers move to the unit circle, transitioning the topological classification of the periodic orbit to Case N7.

Because of the highly irregular shape of 216 Kleopatra, this periodic orbit family provides quite different orbits from the periodic orbits obtained around a massive ring (Broucke & Elipe 2005) or a massive line segment (Palacián et al. 2006). Scheeres (2012) calculated several resonant periodic orbits around 4179 Toutatis. However, the periodic orbit family obtained for 216 Kleopatra is non-resonant. From the orbital positions of this periodic orbit family, one may speculate that the family is generated by the center equilibrium point of 216 Kleopatra. However, the eigenvalues of the center equilibrium point are ±0.001473i, ±0.001175i, and ±0.0005663, which indicates that two families of periodic orbits may be generated by the center equilibrium point. Based on the eigenvalues, one can determine that the respective periods of these two periodic orbit families generated by the center equilibrium point are 1.18475 h and 1.48574 h. As discussed, the continued periodic orbit family considered herein has periods between 2.4257 h and 2.3855 h.

From the results presented above, we can conclude that multiple bifurcations may exist in a periodic orbit family. The Floquet multipliers in the complex plane can rotate 180°, such that the intersection point of the periodic orbit on the Poincaré section rotates 180 °.

## 4. Conclusions

We studied the dynamical mechanism of periodic orbit families about irregular-shaped bodies. The topological classifications of periodic orbits and multiple bifurcations in periodic orbit families have been investigated. We verified the existence of a conserved quantity that can restrict the number of periodic orbits in a fixed energy curved surface. We applied the theoretical results to two irregular-shaped celestial bodies, asteroid 216 Kleopatra and comet 1P/Halley. The gravitational and effective potentials of 216 Kleopatra and 1P/Halley were generated using the polyhedron shape model. The shapes of the Hill regions for 216 Kleopatra presented a bifurcated two-lobe structure while the shapes of the Hill regions for 1P/Halley presented a non-axisymmetric structure.

Several different bifurcation pathways were investigated. We discussed the collision maintaining attribute of Floquet multipliers during the continuation of periodic orbits, which were observed during the continuation of a periodic orbit family about 1P/Halley. Multiple bifurcations were also investigated. When continuing a periodic orbit family about 216 Kleopatra, we observed multiple bifurcations consisting of two real saddle bifurcations and a period-doubling bifurcation. In addition, we also found Floquet multipliers in the complex plane that rotate 180 ° between the two real saddle bifurcations during the continuation of the periodic orbit family.

**Acknowledgements**

This research was supported by the National Science Foundation for Distinguished Young Scholars (Grant No. 11525208), the State Key Laboratory of Astronautic Dynamics Foundation (Grant No. 2016ADL-0202), and the National Natural Science Foundation of China (Grant No. 11372150).

**Appendix**

Table A1. The topological cases for periodic orbits with period-doubling (+1: Floquet multipliers equal +1; −1: Floquet multipliers equal −1; +x: Floquet multipliers on the +x axis except ±1; −x: Floquet multipliers on the −x axis except ±1; UC: Floquet multipliers in the unit circle except ±1; CP: Floquet multipliers in the complex plane except ±1, x axis, and unit circle).

Non-degenerate cases

| Cases | +1 | −1 | +x | −x | UC | CP |
|---|---|---|---|---|---|---|
| N1 | 2 | 0 | 0 | 0 | 0 | 4 |
| N2 | 2 | 0 | 0 | 0 | 4 different | 0 |
| N3 | 2 | 0 | 4 different | 0 | 0 | 0 |
| N4 | 2 | 0 | 2 | 2 | 0 | 0 |
| N5 | 2 | 0 | 0 | 4 different | 0 | 0 |
| N6 | 2 | 0 | 2 | 0 | 2 | 0 |
| N7 | 2 | 0 | 0 | 2 | 2 | 0 |

Degenerate periodic cases

| Cases | +1 | −1 | +x | −x | UC | CP |
|---|---|---|---|---|---|---|
| DP1 | 4 | 0 | 0 | 0 | 2 | 0 |
| DP2 | 4 | 0 | 2 | 0 | 0 | 0 |
| DP3 | 4 | 0 | 0 | 2 | 0 | 0 |
| DP4 | 6 | 0 | 0 | 0 | 0 | 0 |

Krein collision cases

| Cases | +1 | −1 | +x | −x | UC | CP |
|---|---|---|---|---|---|---|
| K1 | 2 | 0 | 0 | 0 | 4, collided | 0 |

Degenerate real saddle cases

| Cases | +1 | −1 | +x | −x | UC | CP |
|---|---|---|---|---|---|---|
| DR1 | 2 | 0 | 4, Degenerate | 0 | 0 | 0 |
| DR2 | 2 | 0 | 0 | 4, Degenerate | 0 | 0 |

Period-doubling cases

| Cases | +1 | −1 | +x | −x | UC | CP |
|---|---|---|---|---|---|---|
| PD1 | 4 | 2 | 0 | 0 | 0 | 0 |
| PD2 | 2 | 4 | 0 | 0 | 0 | 0 |
| PD3 | 2 | 2 | 0 | 0 | 2 | 0 |
| PD4 | 2 | 2 | 2 | 0 | 0 | 0 |
| PD5 | 2 | 2 | 0 | 2 | 0 | 0 |

Table A2. The pathways for bifurcations in periodic orbit families

Period-doubling bifurcations.

| Identifier | Pathways |
|---|---|
| I | Case DP1 → Case PD1 → Case DP3 |
| II | Case PD3 → Case PD2 → Case PD5 |
| III | Case N1 → Case PD2 → Case N5 |
| IV | Case N2 → Case PD3 → Case N7 |
| V | Case N4 → Case PD4 → Case N6 |
| VI | Case N5 → Case PD5 → Case N7 |

Tangent bifurcations

| Identifier | Pathways |
|---|---|
| I | Case N2 → Case DP1 → Case N6 |
| II | Case N3 → Case DP2 → Case N6 |
| III | Case N4 → Case DP3 → Case N7 |
| IV | Case DP1 → Case DP6 → Case DP2 |

Neimark-Sacker bifurcations

| Identifier | Pathways |
|---|---|
| I | Case N1 → Case PK1 → Case N2 |

Real saddle bifurcations

| Identifier | Pathways |
|---|---|
| I | Case N1 → Case DR1 → Case N3 |

# References

Alberti, A., & Vidal, C. 2007, Celest. Mech. Dynam. Astron., 98, 75

Baer, J., Chesley, S. R., & Matson, A. R. D. 2011, Astron. J., 141, 205

Blaikie, A., Saines, A. D., Schmitthenner, M. et al. 2014, PhRvE, 89, 042917

Blesa, F. 2006, Monogr. Semin. Mat. García Galdeano., 33, 67

Broucke, R. A., & Elipe, A. 2005, Regul. Chaotic Dyn., 10, 129

Chanut, T. G. G., Winter, O. C., & Tsuchida, M. 2014, Mon. Not. R. Astron. Soc., 438, 2672

Chanut, T. G. G., Winter, O. C., Amarante, A., & Araújo, N. C. S. 2015a, Mon. Not. R. Astron. Soc., 452. 1316

Chanut, T. G. G., Aljbaae, S., & Carruba, V. 2015b, Mon. Not. R. Astron. Soc., 450, 3742

Chappell, J. M., Chappell, M. J., Iqbal, A., & Abbott, D. 2012, Phys. Int., 3, 50

Descamps, P., Marchis, F., Berthier, J. et al. 2011, Icar, 211, 1022

Fukushima, T. 2010, Celest. Mech. Dynam. Astron., 108, 339

Galán, J., Muñoz-Almaraz, F. J., Freire, E., Doedel, E., Vanderbauwhede, A. 2002, PhRvL, 88, 241101

Greenberg, A. H., & Margot, J. L. 2015, Astron. J., 150, 114

Hirabayashi, M., & Scheeres, D. J. 2015, Astrophys. J. Lett., 798, L8

Jiang, Y., & Baoyin, H. 2014, J. Astrophys. Astron., 35, 17

Jiang, Y., Baoyin, H., Li, J., & Li, H. 2014, Astrophys. Space Sci., 349, 83

Jiang, Y. 2015, Earth, Moon, and Planets, 115, 31

Jiang, Y., Yu, Y., & Baoyin, H. 2015a, Nonlinear Dynam., 81, 119

Jiang, Y., Baoyin, H., & Li, H.: 2015b, Astrophys. Space Sci., 360, 63

Jiang, Y., Baoyin, H., & Li, H. 2015c, Mon. Not. R. Astron. Soc., 452, 3924

Jiang, Y., Baoyin, H., Wang, X., et al. 2016, Nonlinear Dynam., 83, 231

Lindner, J. F., Jacob, L., King, F. W., & Amanda, L. 2010, Phys. Rev. E, 81, 036208

Lu, X., Zhao, H., & You, Z. 2013, Res. Astron. Astrophys., 04, 471

Lu, X., Zhao, H., & You, Z. 2014, Earth Moon Planets, 112, 73

Marsden, J. E., & Ratiu, T. S. 1999, Springer Berlin

Meyer, K. R. 1998, Phys. D, 112, 310

Moore, C. 1993, Phys. Rev. L., 70, 3675

Muñoz-Almaraz, F. J., Freire, E., Galán, J., Doedel, E., Vanderbauwhede, A. 2003, JPhD, 181, 1

Najid, N. E., Elourabi, E. H., & Zegoumou, M. 2011, Res. Astron. Astrophys. 11, 345

Neese, C., Ed. 2004, Small Body Radar Shape Models V2.0. NASA Planetary Data System

Ostro, S. J., Hudson, R. S., & Nolan, M. C. 2000, Science, 288, 836

Palacián, J. F., Yanguas, P., & Gutiérrez-Romero, S. 2006, SIAM J. Appl. Dyn. Syst.,

5, 12

Peale, S. J., & Lissauer, J. J. 1989, Icarus, 79, 396

Riaguas, A., Elipe, A., & Lara, M. 1999, Celest. Mech. Dyn. Astron., 73, 169

Romanov, V. A., & Doedel, E. J. 2012, Int. J. Bifur. Chaos., 22, 1230035

Romanov, V. A., & Doedel, E. J. 2014, Int. J. Bifur. Chaos., 24, 1430012

Sagdeev, R. Z., Szabó, F., Avanesov, G. A., et al. 1986, Natur, 321, 262

Sagdeev, R. Z., Elyasberg, P. E., & Moroz, V. I. 1988, Nature, 331, 240

Scheeres, D. J., Ostro, S. J., Hudson, R. S., Werner, R. A. 1996, Icar, 121, 67

Scheeres, D. J. 2012, Acta. Astronaut., 7, 21

Stooke, P., 2002, Small Body Shape Models. EAR-A-5-DDR-STOOKE-SHAPE-MODELS-V1.0. NASA Planetary Data System

Vasilkova, O. 2005, Astron. Astrophys., 430, 713

Werner, R. 1994, Celest. Mech. Dyn. Astron., 59, 253

Werner, R., & Scheeres, D. J. 1997, Celest. Mech. Dyn. Astron., 65, 313

Yu, Y., & Baoyin, H. 2012, Mon. Not. R. Astron. Soc., 427, 872

Zotos, E. E. 2014, Nonlinear Dynam., 78, 1389

Zotos, E. E. 2015a, Nonlinear Dynam., 82, 357

Zotos, E. E. 2015b, Nonlinear Dynam., 82, 1233